\documentclass{article}

\usepackage{arxiv}

\usepackage[utf8]{inputenc} 
\usepackage[T1]{fontenc}    
\usepackage{hyperref}       
\usepackage{url}            
\usepackage{booktabs}       
\usepackage{amsfonts}       
\usepackage{nicefrac}       
\usepackage{microtype}      
\usepackage{lipsum}
\usepackage{graphicx}
\graphicspath{ {./images/} }

\title{A Cellular Automata Approach to Donation Game}

\author{
  Marcin Kowalik \\
  School of Coumputing and Information\\
  Rzeszów University of Technology\\
  Rzeszów, Poland \\
  \texttt{mkowalik@prz.edu.pl} \\
   \And
  Przemysław Stokłosa \\
  School of Coumputing and Information\\
  Institute of Management and Information Technology\\
  Bielsko-Biała, Poland \\
  \texttt{przemyslaw.stoklosa@gmail.com} \\
  \AND
  Mateusz Grabowski \\
  Rzeszów University of Technology \\
  Rzeszów, Poland  \\
  \texttt{173145@stud.prz.edu.pl} \\
  \And
  Janusz Starzyk \\
  University of Information Technology and Management \\
  Rzeszów, Poland  \\
  Ohio University \\
  Athens, OH, USA\\
  \texttt{starzykj@gmail.com} \\
  \And
  Paweł Raif \\
  Silesian University of Technology \\
  Gliwice, Poland \\
  \texttt{pawel.raif@gmail.com} \\
}

\begin{document}
\maketitle
\begin{abstract}
The donation game is a well-established framework for studying the emergence and evolution of cooperation in multi-agent systems. The cooperative behavior can be influenced by the environmental noise in partially observable settings and by the decision-making strategies of agents, which may incorporate not only reputation but also traits such as generosity and forgiveness. Traditional simulations often assume fully random interactions, where cooperation is tested between randomly selected agent pairs. In this paper, we investigate cooperation dynamics using the concept of Stephen Wolfram’s one-dimensional binary cellular automata. This approach allows us to explore how cooperation evolves when interactions are limited to neighboring agents. We define binary cellular automata rules that conform to the donation game mechanics. Additionally, we introduce models of perceptual and action noise, along with a mutation matrix governing the probabilistic evolution of agent strategies. Our empirical results demonstrate that cooperation is significantly affected by agents’ mobility and their spatial locality on the game board. These findings highlight the importance of distinguishing between entirely random multi-agent systems and those in which agents are more likely to interact with their nearest neighbors.
\end{abstract}


\section{Introduction}
Cooperation among organisms represents a compelling example of emergent behavior, where individuals---whether embodied biological entities or abstract agents---coordinate their actions to achieve mutual or collective benefit. Such cooperation can arise at multiple levels: from interactions between individual agents, to the formation of cohesive social groups, and even inter-species collaborations. Examples provided by nature show us that the intelligence of collaborating entities does not impact the emergence of cooperation. Cooperative behavior has been observed at the micro scale (e.g., bacteria) and the macro scale (e.g., animals, human societies). While cooperation has been extensively studied across disciplines---including biology, economics, and artificial intelligence---it remains a phenomenon that is not yet fully understood~\cite{Trivers1971,axelrod1984evolution, nowak2006five}.

In primate societies, cooperation is often mediated by \textit{reputation systems} and \textit{social norms}, which are key components of \textit{indirect reciprocity}~\cite{1984_Axelrod_Biology_of_moral_systems,Nowak_Sigmund_1998}. These mechanisms motivate individuals to cooperate even in situations where the immediate cost of cooperation is borne by only one party. The long-term benefit may not necessarily return to the cooperator directly, but instead strengthens social cohesion, contributes to group stability, or supports the survival of the broader population.

A widely used approach to studying cooperation is \textit{agent-based simulation}, where artificial agents engage in strategic interactions such as the \textit{Prisoner’s Dilemma} or the \textit{Donation Game}~\cite{Nowak_Sigmund_1998, santos2005scale, sigmund2010calculus, Schmid_2021}. In this paper, we focus on the Donation Game as a framework for modeling the emergence and stability of cooperation in a population of interacting agents.

\begin{figure}[ht]
\centering
\includegraphics[width=0.75\linewidth]{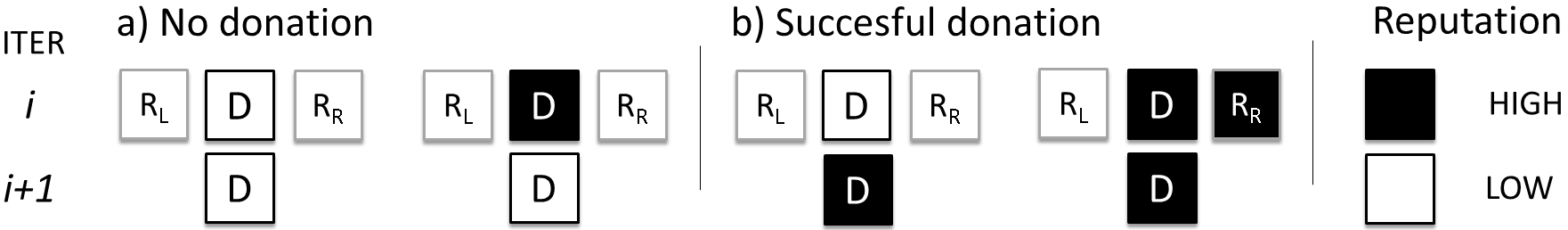}

\caption{(a) If the donor cell \textit{D} does not donate to either of its neighboring recipient cells {$R_{L}$} or {$R_{R}$} during iteration \textit{i}, it transitions to a low-reputation state (white) in iteration \textit{i+1}. (b) When a donation occurs, the donor cell \textit{D} gains a high-reputation state (black) in the next iteration. The reputation (color) of recipient cells  {$R_{L}$} and {$R_{R}$} remains unchanged regardless of whether they receive a donation.}
\label{fig:Donation_no_donation}
\end{figure}

\begin{figure}[p]
\centering
\includegraphics[width=0.65\linewidth]{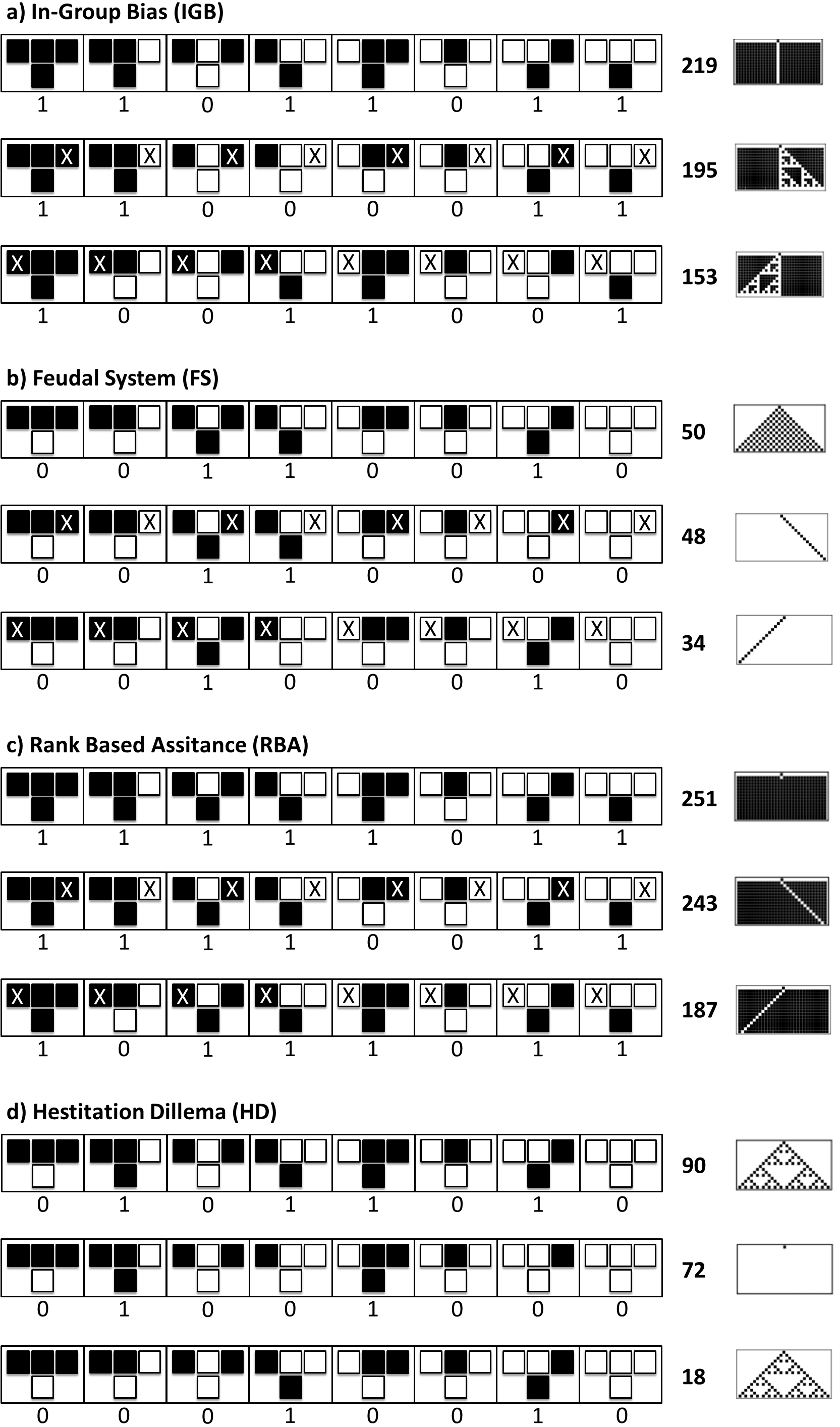}

\caption{Wolfram cellular automata rules \cite{wolfram1983statistical, Wolfram2002} aligned with variants of Donation Game dynamics. (A) In-Group Bias, (B) Feudal System, (C) Rank-Based Assistance, and (D) Hesitation Dilemma. Each row represents a cellular automaton rule whose update behavior reflects the decision logic of the corresponding Donation Game variant. Black and white cells denote high and low reputations, respectively, while 'X' marks indicate cells that are ignored by the donor during decision-making. On the right, the corresponding Wolfram rule number is shown alongside the space-time pattern generated from a single high-reputation seed cell placed at the center of the first row (iteration 0).
}
\label{fig:Rules_explained}
\end{figure}


Cellular automata (CA) are simple computational models composed of a grid of cells, where each cell updates its state based on local rules and the states of its neighbors. John von Neumann introduced CA in the 1950s as a model of self-replication~\cite{vonneumannTheorySelfReproducingAutomata1966}. They have since become a foundational tool for studying complex systems composed of many interacting components, and were further popularized in the 1980s by Stephen Wolfram~\cite{wolfram1983statistical, Wolfram2002}, who investigated the emergence of complex behavior from simple rule sets.

The integration of CA with game-theoretic concepts—particularly in the study of cooperation—was pioneered by Nowak and May~\cite{nowakEvolutionaryGamesSpatial1992}, who demonstrated how spatial structure and local interactions influence the evolution of cooperative strategies. This paper adopts the CA framework to study the Donation Game, modeling agents as binary-state cells governed by socially interpretable cooperation rules.


\section{Donation Game \&  1D  Celullar binary Automata}
Typically, the donation game is a form of the prisoner's dilemma where agents can either donate or not. A donation provides a benefit \textit{b} to the recipient \textit{R} at a cost \textit{c} to the donor \textit{D}. Usually \textit{c} is smaller than \textit{b}. Agents' strategies could be cooperative (always donate), selfish (never donate), or conditional (e.g., tit-for-tat).

In the binary 1D cellular automata (BCA) donation game, agents are represented as cells in a one-dimensional grid (a row or vector), where each cell can exist in one of two possible states: low reputation (white square) or high reputation (black square). The actions of an agent depend solely on the states of its nearest neighbors, following predefined cooperation rules that dictate when donations occur. The donor agent, as illustrated in Figure \ref{fig:Donation_no_donation}a, is represented by the central cell; Neighboring cells are potential recipients. A successful donation increases the donor's reputation, turning a previously white (low-reputation) cell black (high-reputation) or maintaining the black state for an additional iteration. Conversely, a failure to donate results in the donor's reputation degrading to low (white). Notably, the recipient cell does not change its reputation as a result of the transaction (Fig. ~\ref{fig:Donation_no_donation}b). Binary reputation systems have also been explored in prior work \cite{Schmid_2021, Schmid_2021nature, Ohtsuki_2006}, though not in the context of BCA.

\subsection{Rules of donation}

For 1D BCA, we identify three distinct cooperation rule sets that shape the dynamics of donation behavior:
\begin{enumerate}
    \item \textbf{In-Group Bias (IGB)}: Donation occurs exclusively between agents of the same reputation. This models preferential cooperation within social groups, where agents assist only those of similar status (Fig.  \ref{fig:Rules_explained}a).

    \item \textbf{Feudal System (FS)}: Donations are unidirectional, flowing only from low-reputation donors to high-reputation recipients. This reinforces a hierarchical structure where resources transfer upward, and high-status agents do not reciprocate (Fig.  \ref{fig:Rules_explained}b).

    \item \textbf{Rank-Based Assistance (RBA)}: Donations are allowed if the recipient has either the same or a higher reputation than the donor, but never a lower one. This model represents a meritocratic system where cooperation is based on rank rather than strict reciprocity (Fig.  \ref{fig:Rules_explained}c).
\end{enumerate}

In this paper, we adopt Wolfram’s classification scheme for binary cellular automata to characterize the emergent cooperation patterns within these rule sets~\cite{wolfram1983statistical,Wolfram2002}. Figure \ref{fig:Rules_explained} presents the corresponding rule numbers from Wolfram’s taxonomy and the spatiotemporal patterns generated when each rule is initialized with a single high-reputation (black) cell at the center of the first row.

The \textbf{In-Group Bias} behavior is exemplified by rule 219, which permits donation if at least one of the two neighboring cells of the donor has the same reputation as the donor itself. While this rule inherently favors cooperation within groups, it surprisingly does not hinder overall cooperation. Instead, it fosters widespread cooperation, ultimately leading to a state where nearly all cells become donors (Fig.  \ref{fig:Rules_explained}a). If donations are restricted to a single designated neighbor (either left or right), variations of this rule emerge: rule 195 governs donation to left-neighbor recipients, while rule 153 applies to right-neighbor recipients (Fig.  \ref{fig:Rules_explained}a).

The \textbf{Feudal System} behavior, where donations occur exclusively to high-reputation cells, is governed by rule 50 (Fig.  \ref{fig:Rules_explained}b). This rule results in a checkerboard pattern, where equal numbers of cells exist in low- and high-reputation states. When donation is restricted to a specific neighbor, distinct rules arise: rule 48 applies to left-side recipients, while rule 34 applies to right-side recipients (Fig.  ~\ref{fig:Rules_explained}b). These descendant rules exhibit significantly reduced cooperation compared to rule 50, as restricting donation pathways limits the spread of high-reputation states.

The \textbf{Rank-Based Assistance} model, which permits donation to equal- or higher-status recipients, is best described by rule 251 (Fig.  ~\ref{fig:Rules_explained}c). This rule promotes widespread cooperation, with agents acting as donors in nearly all neighborhood configurations—except when both neighbors hold low reputations. Among the twelve studied rules, rule 251 promotes cooperation most strongly. Variants of this rule include rule 187 and rule 243 (Fig.  \ref{fig:Rules_explained}c), where donation bias is applied toward left or right recipients, respectively. These descendant rules exhibit slightly lower levels of cooperation compared to rule 251 but still maintain a high degree of agent participation in donation exchanges.

In the classical Donation Game, a donor interacts with a single recipient~\cite{Nowak_Sigmund_1998}. However, when modeled using 1D CA, a donor may have one or two nearest neighbors as potential recipients. Introducing local neighborhoods into the Donation Game framework gives rise to the concept of \textbf{donor hesitation}, which occurs when a donor faces multiple eligible recipients but must choose only one to support. In cases where the donor cannot decide between two equally eligible neighbors, this indecision may lead to inaction, resulting in no donation during that interaction. The \textbf{hesitation dilemma (HD)}  can be easily incorporated into the cellular automaton model through a specially selected rule set.

In the context of  \textbf{In-Group Bias} behavior, hesitation in cases where both neighboring recipients belong to the same class is captured by rule 90 (Fig.  \ref{fig:Rules_explained}d). This rule also effectively models \textbf{Rank-Based Assistance} scenarios that include hesitation, where a donor consistently prefers to support only the high-reputation or only the low-reputation neighbor when faced with one high and one low-reputation recipient. Conversely, when a low-reputation donor avoids donating to a higher-reputation recipient and is similarly unwilling to choose a peer of equal low status, this form of hesitation is captured by rule 72 (Fig.  \ref{fig:Rules_explained}d) - the most restrictive rule among those analyzed, as it most strongly inhibits donation. Within the \textbf{Feudal System}, where donations occur exclusively from low-to-high-reputation agents, hesitation arises in a single specific configuration and is best represented by rule 18 (Fig.  \ref{fig:Rules_explained}d).

\begin{figure}[ht]
\centering
\includegraphics[width=0.65\linewidth]{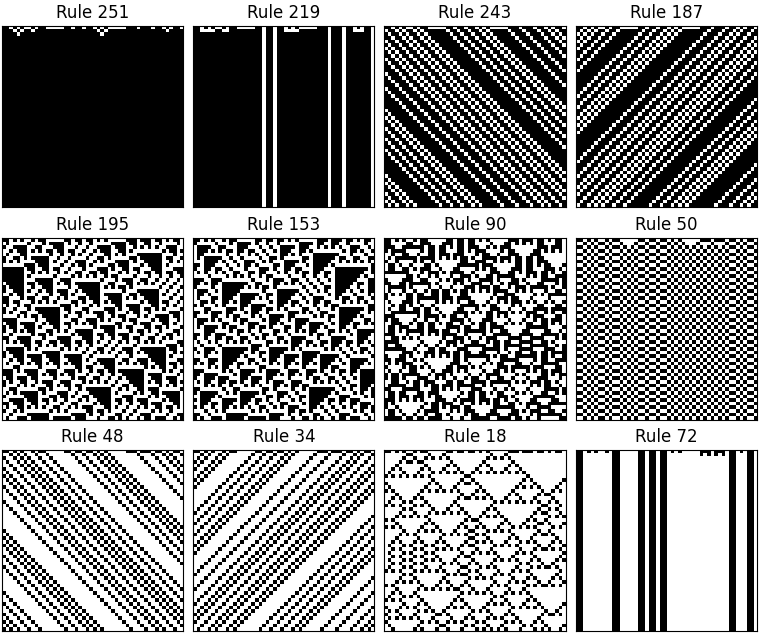}

\caption{Patterns generated by different cellular automata rules, starting from an initial random configuration in the first row. Each grid consists of 50 cells evolving over 50 time steps. Black cells represent high-reputation states, while white cells represent low-reputation states. The observed patterns illustrate how different cooperation rules influence the evolution of reputation dynamics over time.}
\label{fig:Rules_start_from_random_pattern}
\end{figure}

The first experiment involved executing each of the selected rules on a randomly generated binary pattern of 50 cells, evolving it over 50 iterations under periodic boundary conditions, which were applied consistently across all experiments. The results, presented in (Fig.  \ref{fig:Rules_start_from_random_pattern}), show that the resulting spatiotemporal patterns can be broadly categorized as either periodic or aperiodic with chaotic dynamics. In particular, certain rules generate local spatiotemporal structures; for example, rules 153 and 18 create triangular "pockets" of persistently high‑ and low‑reputation cells, respectively. Alternatively, the patterns can be classified based on the proportion of black cells, representing agents in a high-reputation state.

Interestingly, aside from the purely altruistic rule—defined as "always donate" (rule 255), which is not illustrated here due to its trivial outcome of an entirely black pattern—the \textbf{RBA} rules (251, 243, and 187), which follow the principle "do not donate to agents of lower status," resulted in populations where the majority of agents attained high reputation. A particularly surprising result was observed for \textbf{IGB} (rule 219), which enforces donations only between members of the same class; this rule led to a population in which nearly every agent achieved high reputation. In contrast, the opposite effect was seen in rule 72 (\textbf{RBA with HD}), where only a small fraction of agents retained high reputation throughout the system's evolution. The remaining rules—50, 48, and 34 (\textbf{FS}), rule 18 (\textbf{FS with HD}), and rule 90 (applied to \textbf{IGB} or \textbf{RBA, both with HD})—produced populations in which at least 50\% of agents remained in a low-reputation state (see Figure \ref{fig:Rules_start_from_random_pattern} for all patterns).

\begin{figure}[ht]
\centering
\includegraphics[width=0.55\linewidth]{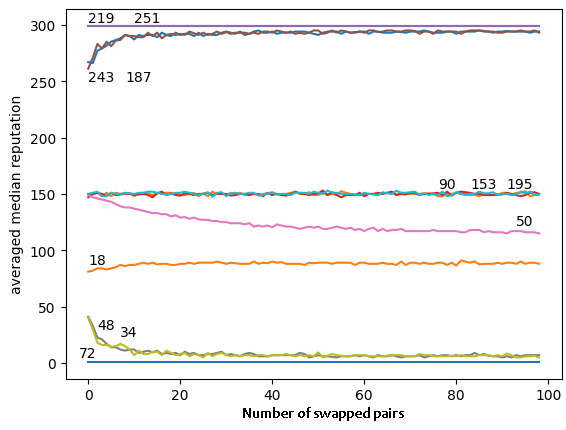}

\caption{Averaged median reputation in a 100-agent system governed by different cellular automata rules, plotted against the number of swapped agent pairs. The system was evolved over 300 iterations under periodic boundary conditions. Numbers shown on the plot indicate the rule numbers associated with the nearest curves.}
\label{fig:median_reputation_vs_shuffling}
\end{figure}

\subsection{Agent mobility - random swapping}
The 1D CA version of the Donation Game assumes that agents remain fixed in their positions. However, agents may move across the game board in more realistic scenarios, interacting with different agents over time. We introduced agent mobility by randomly swapping a fixed number of agent pairs between consecutive iterations to investigate this factor. On a 100‑cell grid, we varied the swap rate from 0 to 100 pair swaps; swapping 100 pairs per step yields a rough approximation of a randomized configuration\cite{1981_Diaconis, 2000_Trefethen}. 

In the BCA framework, an agent’s donation activity is equivalent to the sum of black cells in a column. Dividing this sum by the number of steps (rows) provides the percentage of iterations in which an agent maintained a high reputation. Summing across rows yields the number of donations that occurred in a given iteration. Figure \ref{fig:median_reputation_vs_shuffling}  shows the average median reputation for a 100‑cell system run for 300 iterations under the different swap rates.

The results indicate that \textbf{In-Group Bias} rules (219, 153, 195, and 90) remain largely unaffected by random neighbor swaps. Among the \textbf{Rank-Based Assistance} rules, 187 and 243 show an increase in median reputation, which saturates with a small number of swaps, while rules 251 and 90 remain unchanged. In contrast, random swapping completely destroys cooperation in rule 72, where hesitation prevents donations. Within the \textbf{Feudal System}, rules 50, 48, and 34 show a decline in donations as swapping increases, but when hesitation is introduced (rule 18), donation rates slightly increase before stabilizing. Rules 251 and 219, which inherently encourage donations, result in all cells donating and maintaining a high reputation across all iterations. Conversely, rule 72 inhibits donations, leading to an overall low-reputation population. This demonstrates that agent cooperation is influenced by the rules governing interactions and the probability of engaging with different agents over time.

\begin{figure}[p]
\centering
\includegraphics[width=0.7\linewidth]{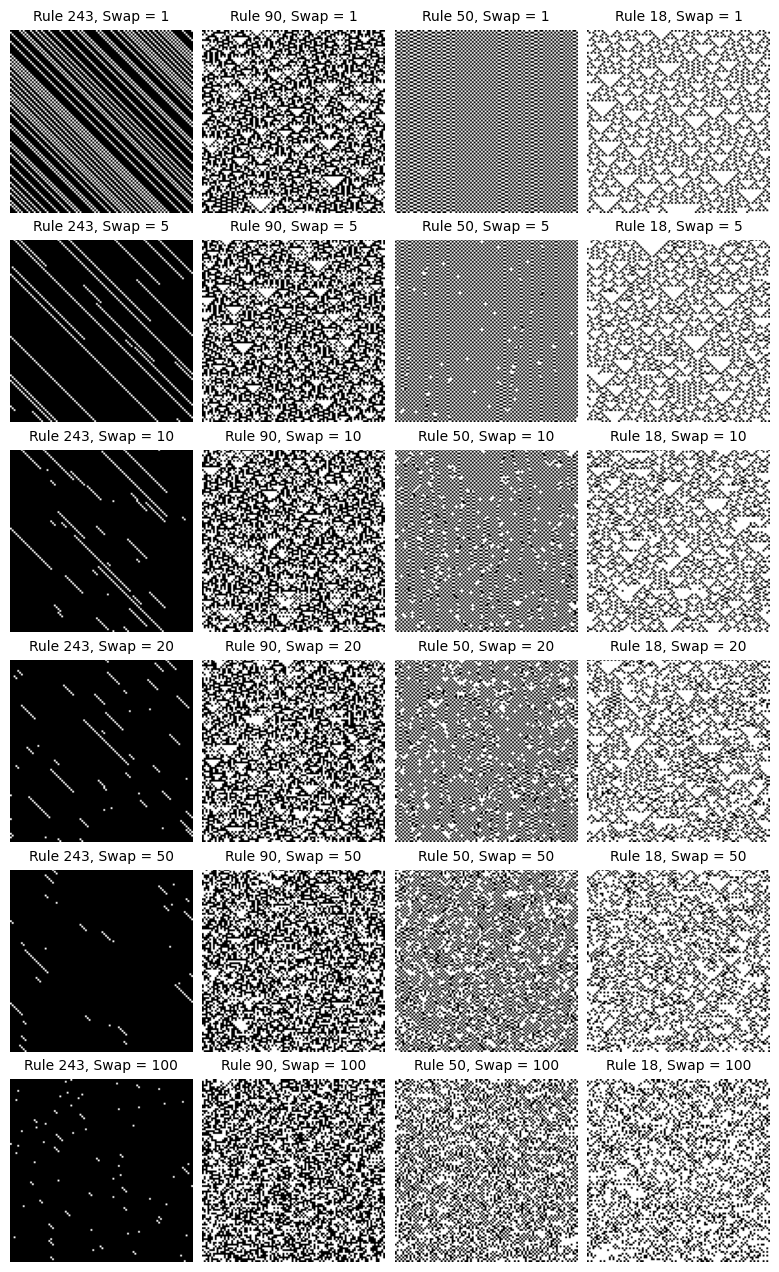}

\caption{Spatiotemporal evolution of agent reputations over 100 iterations for selected rules under varying levels of agent mobility introduced through random swapping. Each subplot is a space-time diagram: columns represent fixed grid positions (i.e., specific cell locations), and rows represent successive time steps. Black and white pixels indicate high and low reputation states, respectively. The four columns correspond to Rules 243, 90, 50, and 18; each row of subplots shows results at increasing swap levels ( 1, 5, 10, 20, 50 and 100 pairs swapped per iteration). As agent mobility increases, structured patterns become increasingly disrupted. For small swap values, particularly under Rules 50 and 18, structured and novel spatiotemporal patterns emerge, which are lost at higher mobility levels.}
\label{fig:Rules_vs_shuffle}
\end{figure}

Figure \ref{fig:Rules_vs_shuffle} shows the spatiotemporal patterns of grid cells—that is, the reputation histories of the locations themselves—generated under four representative rules (243, 90, 50, 18) and six swap rates (1, 5, 10, 20, 50, 100 pair swaps per iteration). In this grid‑cell view, each column corresponds to a fixed position (site) on the lattice rather than to a particular agent; a given cell may therefore be occupied by different agents at different time steps. By contrast, an agent‑centric view—tracking individual agents as they move—would obscure these spatiotemporal dependencies, because constantly swapping agents reassigns them to new locations.

At low swap rates (e.g., five pair swaps per iteration), new types of correlations emerge between neighboring sites. This is especially evident for Rule 50. In the absence of swapping, rule 50 produces a chessboard pattern in which lattice sites alternate between high and low reputation every two iterations (see Figure \ref{fig:Rules_explained}). Introducing even modest mobility generates novel low‑reputation spatiotemporal structures like white crosses, white triangles, and vertically and horizontally synchronized pairs of neighboring sites that drift left or right over time. When two synchronized pairs are separated by a single site, a white V‑shaped structure may appear to re‑establish the chessboard pattern. At higher swap rates, other, additional low or high reputation structures emerge, such as double white crosses or diagonal black or white lines (see the pattern for rule 50 at swap = 1 and 5 in Figure \ref{fig:Rules_vs_shuffle}). For all rules, recognizable spatiotemporal structures persist up to a swap rate of about 20; beyond that point, increasing randomness progressively erases local spatiotemporal order.

\subsection{Directed movement strategy}
Unlike random swaps, which relocate agents unpredictably, directed movement follows a fixed trajectory. To evaluate its effect, we shifted every even‑indexed agent two cells to the right at each iteration (e.g., an agent in cell 1 moved to cell 3, an agent in cell 3 to cell 5, and so on). Combining deterministic shift with varying levels of random swapping shows only minor changes in the average median reputation. Because the resulting curves were similar to those obtained with swapping alone (Fig.  \ref{fig:median_reputation_vs_shuffling}), we do not present the additional plots.

The differences we did observe were modest: rules 243 and 187 showed slightly higher median reputations when the swap rate was zero, whereas rules 48 and 34 exhibited marginally lower values. The largest deviation occurred under rule 50, where the initial median reputation was roughly 140 and declined more slowly with increasing swap rates than in Figure \ref{fig:median_reputation_vs_shuffling}—yet even this effect was too small to warrant a separate figure.

\begin{figure}[p]
\centering
\includegraphics[width=0.7\linewidth]{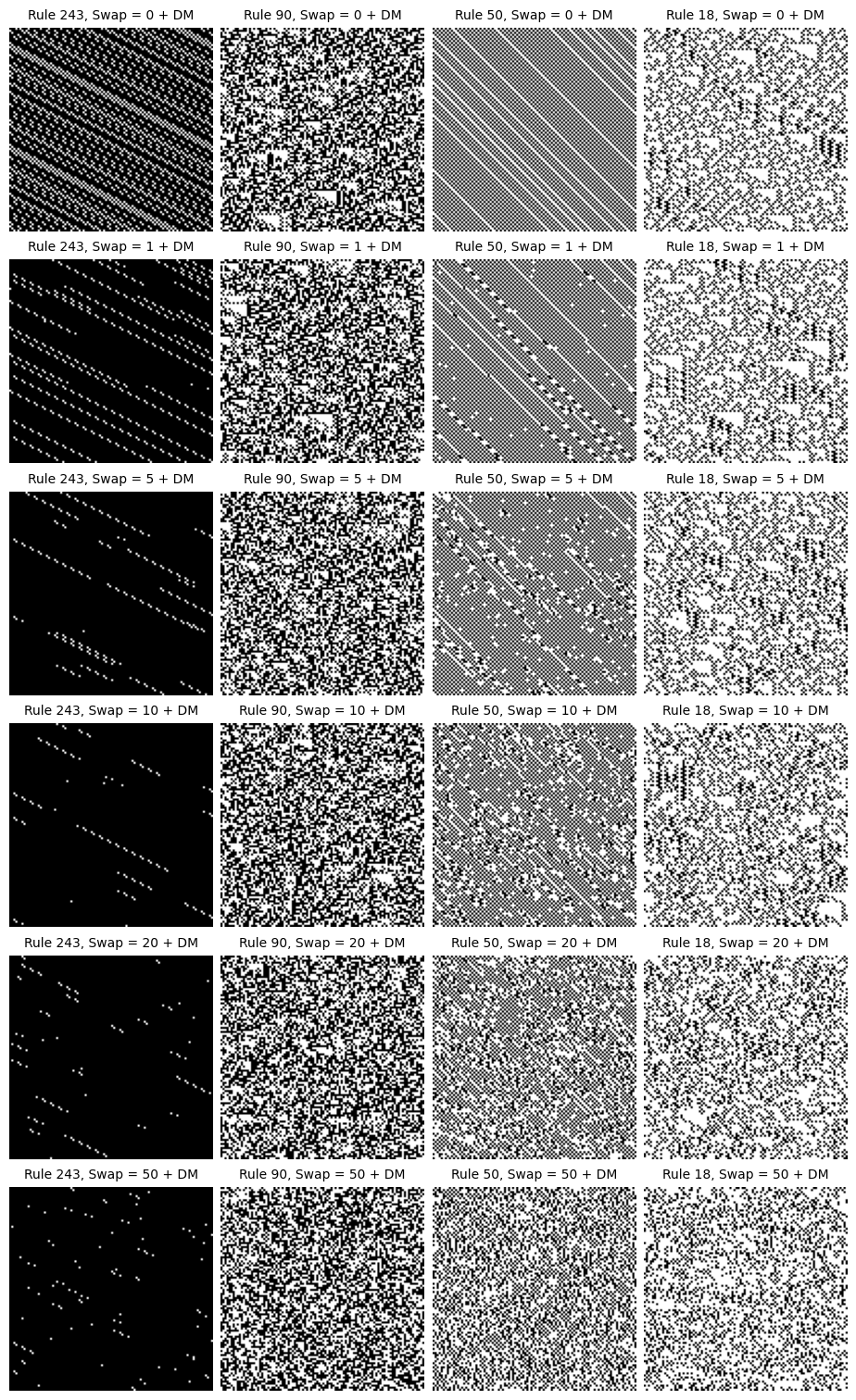}

\caption{Spatiotemporal evolution of agent reputations over 100 iterations for selected rules under varying levels of agent mobility introduced through random swapping and directed motion. Each subplot is a space-time diagram: columns represent fixed grid positions (i.e., specific cell locations), and rows represent successive time steps. Black and white pixels indicate high and low reputation states, respectively. The four columns correspond to Rules 243, 90, 50, and 18; each row of subplots shows results at increasing swap levels (0, 1, 5, 10, 20, and 50 pairs swapped per iteration). As agent mobility increases, structured patterns become increasingly disrupted. For small swap values, particularly under Rules 50 and 18, structured and novel spatiotemporal patterns emerge, which are lost at higher mobility levels.}
\label{fig:Rules_vs_shuffle_walking}
\end{figure}

A sharper contrast emerges in the grid‑site visualizations (Fig.  \ref{fig:Rules_vs_shuffle_walking}). Under rule 50, directed shifting produces a population of high‑ and low‑reputation agents that migrate diagonally across the lattice. Some low‑reputation agents travel in pairs, appearing as white streaks that cut through the black‑and‑white chessboard pattern. Introducing minimal random swapping (1 pair swap per iteration) enlarges certain low‑reputation pairs into quartets; every three steps, however, the central pair of each quartet briefly regains high reputation for two iterations, creating the regularly spaced, thick dashed lines visible in Fig. 
\ref{fig:Rules_vs_shuffle_walking}. Limited swapping also spawns small, cross‑shaped clusters of low reputation and allows the low‑reputation state of one pair member to transfer to a neighboring site, producing intermittent “breaks” in the background pattern. As the swap rate rises, the diagonal lines fragment and the number of white crosses increases, while the overall pattern becomes gradually more random.

A comparable spatio-temporal transfer of reputation is observed under rule 18. Directed shifting produces vertical, black zig‑zag columns that persist until the swap rate reaches about 20. At this threshold, the zig‑zag pattern survives only four iterations, whereas at lower swap rates, it can endure for a dozen or more. The spatiotemporal plot for rule 18 also features white, skewed triangles—local pockets of low‑reputation agents whose duration ranges from a single iteration to several.

These findings underscore that both the mode of agent movement—random versus directed—and the specific rule governing interactions exert a pronounced impact on the spatiotemporal distribution of high‑ and low‑reputation populations. Furthermore, when movement is combined with even modest noise, localized spatio-temporal clusters of cooperative behaviour can emerge.


\subsection{Reputation and Action Perception under Noise}

The most basic formulation of the BCA framework for modeling the Donation Game assumes a noise-free environment in which agents always perceive the reputations of others accurately, and all actions succeed and are fully observable by every agent. This assumption of full observability aligns with the model proposed by Nowak and Sigmund \cite{Nowak_Sigmund_1998}. However, unlike Nowak and Sigmund's approach, which relies on randomly selected agent pairs, the BCA framework introduces spatial proximity as a defining feature: agents interact primarily with their immediate neighbors. This spatial constraint reflects real-world settings, where individuals are more likely to engage with those physically or socially closest to them.

However, in more realistic settings, agent behavior is influenced not only by spatial proximity but also by perceptual limitations and environmental uncertainty. One such factor is imperfect perception: agents may fail to observe the environment correctly. For example, a donor agent $D_{i}$ may, with a small probability, misperceive the reputation of a recipient $R_{j}$. This perception noise can lead the donor to act contrary to the prescribed rule, either by donating when it is not appropriate or by withholding a donation when it should occur. Additionally, agents may misinterpret the outcomes of others' actions, introducing what we refer to as action noise. In such cases, a successful donation might be misperceived, leading to the donor being unjustly assigned a low reputation. Conversely, unsuccessful donations are consistently interpreted as such, in line with the assumptions in Griffiths and Oren's work \cite{2023_Griffiths_Oren_Generosity}.

Perceptual and feedback errors are commonly observed in natural systems. A well-known example is mimicry in the animal kingdom, where non-poisonous species evolve to resemble toxic ones to deter predators—a form of deception analogous to perceptual noise.

Our BCA model can be easily extended to incorporate both forms of noise:
\begin{itemize}
    \item \textbf{Perception Noise} (Reputation Misjudgment):
    With probability $e_R$, a donor misjudges the recipient’s reputation - perceiving a high-reputation agent as low-reputation and vice versa.

    \item \textbf{Action Noise} (Feedback Error):
    With probability $e_A$, a successful donation is misperceived by observing agents in the next iteration. Despite cooperative behavior, the donor’s reputation may be incorrectly updated as low.
\end{itemize}

\begin{figure}
    \centering
    \includegraphics[width=0.65\linewidth]{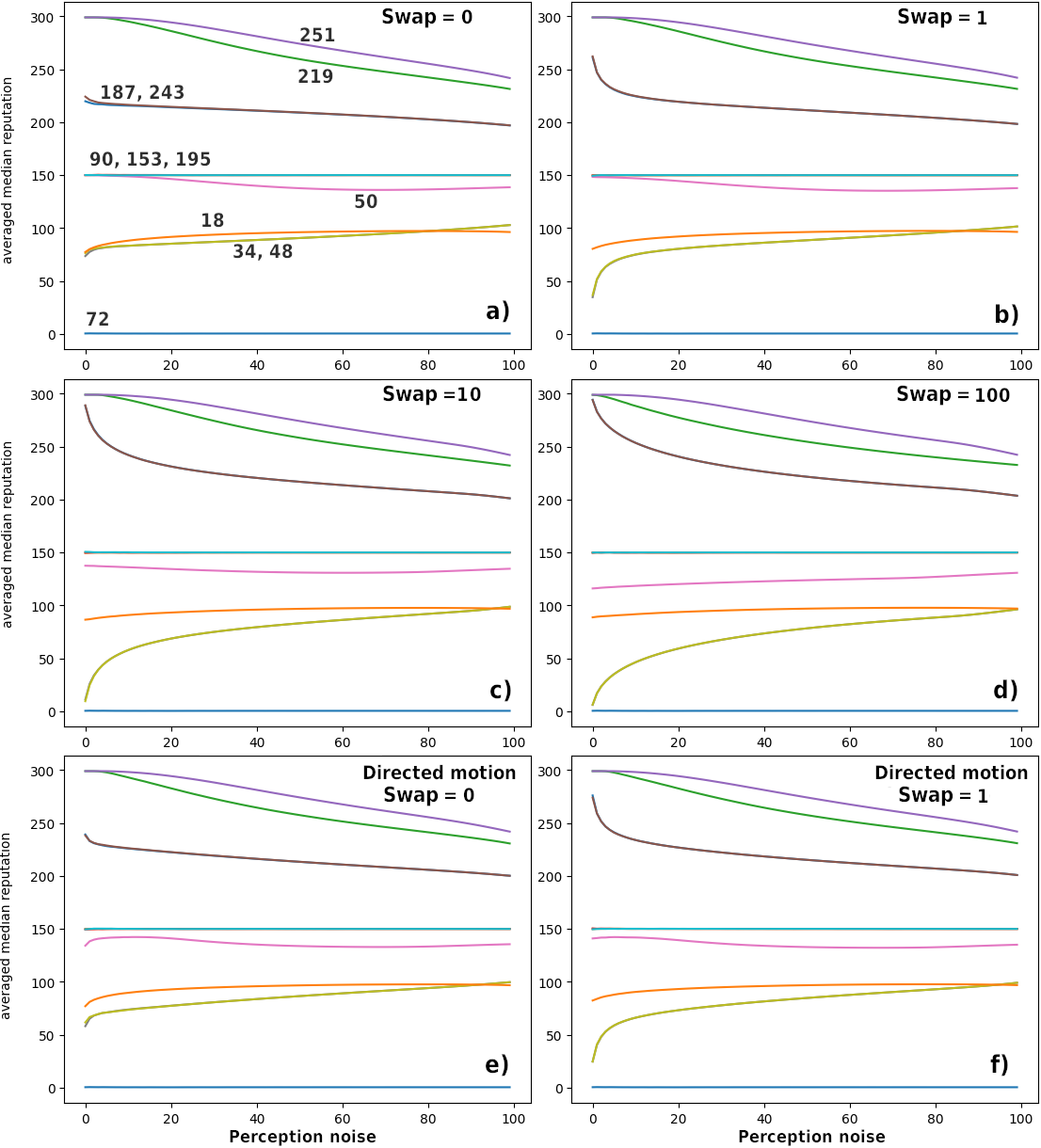}
    
    \caption{Impact of perception noise $e_R$ on the averaged median reputation of agent across behavioral rules and selected swapping level. Subfigures Subfigures (e–f) show results with directed motion enabled.}
    \label{fig:perception_noise}
\end{figure}

\subsubsection{Impact of perception noise}

Figure \ref{fig:perception_noise}a-d shows how perception noise, ranging from 0 to 100\% and selected values (0, 1, 10 and 100) of swap parameter, affects the average median reputation of agent across the full Donation game rule set. Depending on the specific rule, perception noise can either lower or increase this measure. Notably, four rules—90, 153, 195, and 72—are entirely unaffected by misperception.

Rules that strongly promote cooperation, such as 251, 219, 243, and 187, are negatively impacted by perception noise, as uncertainty disrupts their capacity to maintain high reputation through consistent donation behaviour. Interestingly, misperception also helps differentiate between rules 251 and 219, which otherwise yield similar dynamics under noise-free conditions.

By contrast, rules less effective at fostering donations—such as 18, 48, and 34—show a positive response to misperception. Among all rules, rule 50 displays a distinctive two-phase response: its average median reputation initially declines as perception noise increases, but recovers slightly beyond 60\% noise. Rule 18 also exhibits a unique behavior—it is the only rule that marginally maximizes reputation at approximately 70\% perception noise.

When perception noise is combined with agent swapping (Fig. \ref{fig:perception_noise}b-d), the most pronounced effects are seen in rules 72, 18, 243, and 187, which typically produce diagonal line structures in spatiotemporal plots (Fig. \ref{fig:Rules_start_from_random_pattern}). As the swap rate increases, the starting average median reputation tends to rise for cooperation-promoting rules and decline for those that suppress donations. The most significant shifts in these four rules occur within the first 10\% of added perception noise. Beyond a swap rate of about 20, rule 50 transitions from a decrease–increase pattern to a consistently increasing trend. 

Among all rules, the shape of the average median reputation curve for Rule 50 is most sensitive to directed motion (Fig \ref{fig:perception_noise}e-f). When swapping is disabled or minimal (up to two pair swaps per iteration), Rule 50 shows improved cooperation up to about 20\% perception noise. The effect of directed movement is negligible for most other rules. However, some subtle differences can still be observed—for instance, the gap between rules 251 and 219 slightly widens in the presence of directed movement.

\begin{figure}
    \centering
    \includegraphics[width=0.65\linewidth]{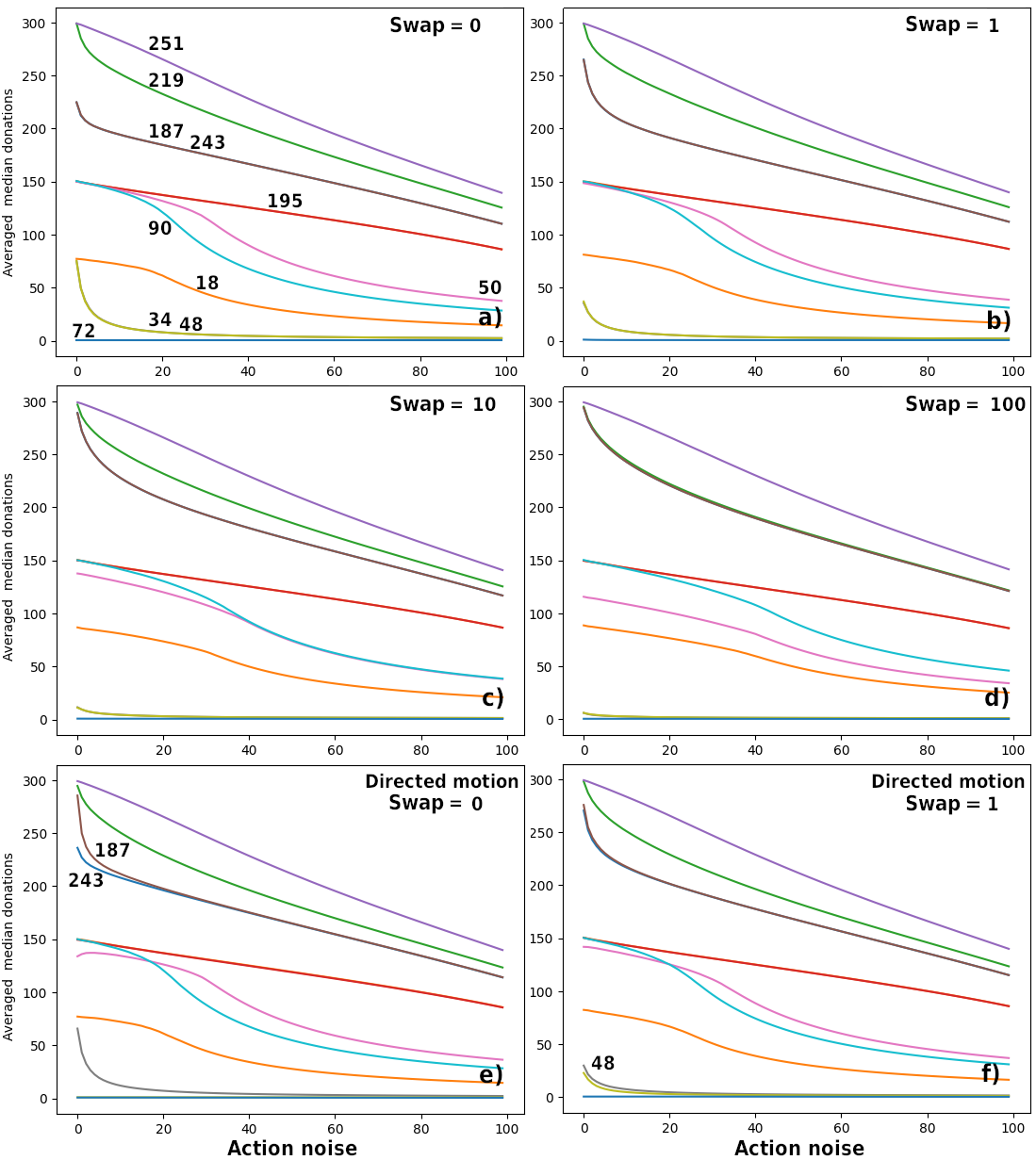}
    \caption{Impact of action noise $e_A$ on the averaged median donations received per agent across behavioral rules and selected swapping levels. Subfigures (e–f) show results with directed motion enabled.}
    \label{fig:action_noise}
\end{figure}

\subsubsection{Impact of action noise}

Figure \ref{fig:action_noise}a-d shows how action noise, ranging from 0 to 100\% and selected values 0, 1, 10, and 100 of the swapping parameter, affects the averaged median donations of an agent, a direct measure of cooperative behavior, across the full Donation game rule set.

In BCA environments free from action noise (or with only perception noise), this donation-based metric closely aligns with the averaged median reputation. However, in the presence of action noise, agents may incorrectly perceive the outcomes of others' donations. As a result, reputation scores no longer reliably reflect actual behavior, limiting their interpretability.

Fig. \ref{fig:action_noise}a  shows that action noise consistently undermines cooperation across nearly all rules, except for the typically non-cooperative Rule 72. The severity of this negative effect increases with the probability of misperceiving donation outcomes. However, the impact of action noise is rule-dependent and non-linear—some rules (219, 187, 243, 34, and 48) exhibit sharp declines in cooperation at low noise levels, while others (50, 90, and 18) display a delayed and broad drop. In contrast, rules such as 251 and 195 degrade more gradually as noise increases.

When swapping is introduced (Fig. \ref{fig:action_noise}b-d), rules 187, 243, and 18 exhibit an increase in donation levels under moderate noise. In contrast, rules 34, 48, and 50 show a decline. Rule 251—the most cooperation-promoting rule—remains largely unaffected by swapping across all noise levels.

Directed motion positively influences cooperation for rules 187 and 243, and negatively for rules 34, 48, and 50 (Fig. \ref{fig:action_noise}e-f). Interestingly, when swapping is disabled, Rule 50—which is otherwise negatively affected by directed motion—shows a slight increase in the number of received donations under low levels of action noise (up to approximately ~3\%). Directed motion also produces asymmetric effects: it enhances cooperation in Rule 187 but reduces it in Rule 243, the mirrored counterpart of Rule 187. At high swap rates (above 50), the impact of directed motion becomes negligible, and we therefore omit those plots for clarity.

\subsection{Evolution}
\begin{figure}
    \centering
     \includegraphics[width=0.65\linewidth]{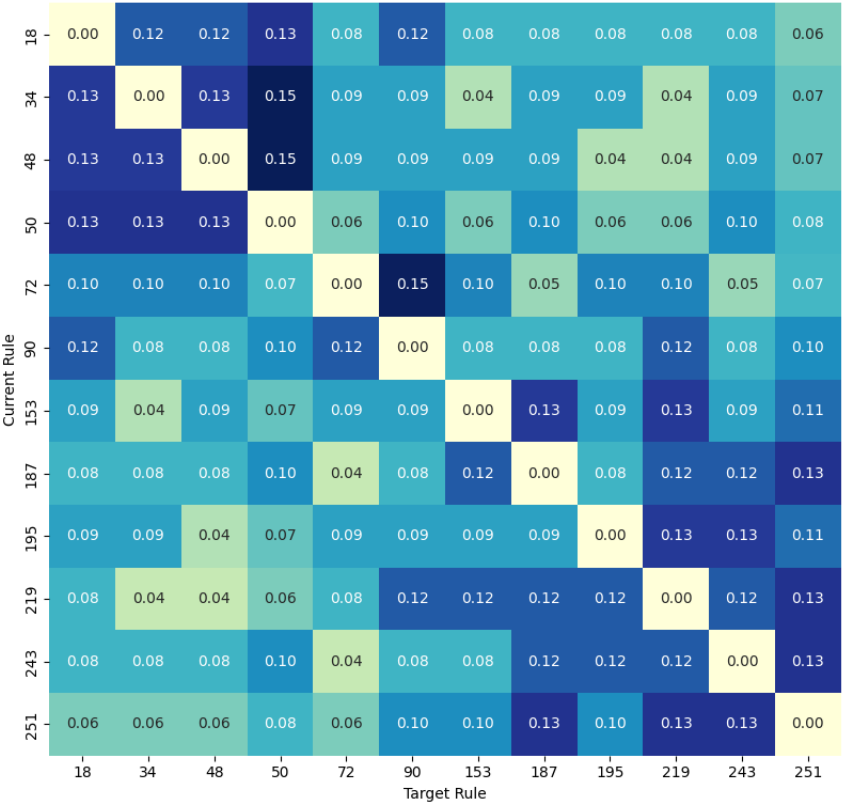}
     \caption{Mutation probability matrix based on bit difference between rules.}
     \label{fig:mutation_matrix}
 \end{figure}

Papers by Nowak and Griffiths\cite{Nowak_Sigmund_1998, 2023_Griffiths_Oren_Generosity} have explored the long-term evolution of indirect reciprocity under mutation and selection, where agent strategies are determined by nuanced reputation metrics (e.g., image scores). In such models, cooperative agents may choose to donate even when recipients possess low reputational standing. In contrast, within the Binary Cellular Automata framework, an agent's strategy is represented by a fixed behavioral rule rather than a continuous assessment of others' reputations.

To enable the adaptive selection of cooperation strategies within a genetically driven BCA model, we introduce a mutation mechanism over a curated set of 12 binary rules. Each rule corresponds to a socially interpretable strategy relevant to the Donation Game, such as In-Group Bias, Rank-Based Assistance, or Feudal Systems. Transitions between strategies are guided by the bitwise Hamming distance between rule encodings. The number of differing bits, $d$ (ranging from 0 to 8), is converted into a similarity score of $1 -\frac{d}{8}$, which is then row-normalized to yield a mutation probability matrix (Fig.~\ref{fig:mutation_matrix}).

We define mutation as a small probability $p_m$ that an offspring adopts a strategy different from that of its parent. When a mutation occurs, the new strategy is selected based on the weighted probabilities from the mutation matrix, which favors transitions to structurally similar rules. Although this setup does not support open-ended rule evolution, it enables adaptive exploration within a fixed strategy space.\footnote{Open-ended evolution allows for the emergence of unbounded behavioral complexity, analogous to biological systems~\cite{nowak2005evolution}.}

To reflect real-world dynamics of indirect reciprocity~\cite{nowak2005evolution}, selection pressure is applied to favor strategies that accumulate the highest number of donations in a heterogeneous rule environment. Accordingly, at the end of each generation, agents produce offspring in proportion to their fitness, measured by the number of donations received. This approach aligns with the mechanics of BCA and eliminates the need to explicitly model donation costs. In cases where both neighboring agents are eligible recipients and the donor's rule does not include the Hesitation Dilemma, the donation is split equally between them. If only one nearest neighbor qualifies, that agent receives the full donation of value 1.

\begin{figure}
    \centering
    \includegraphics[width=0.75\linewidth]{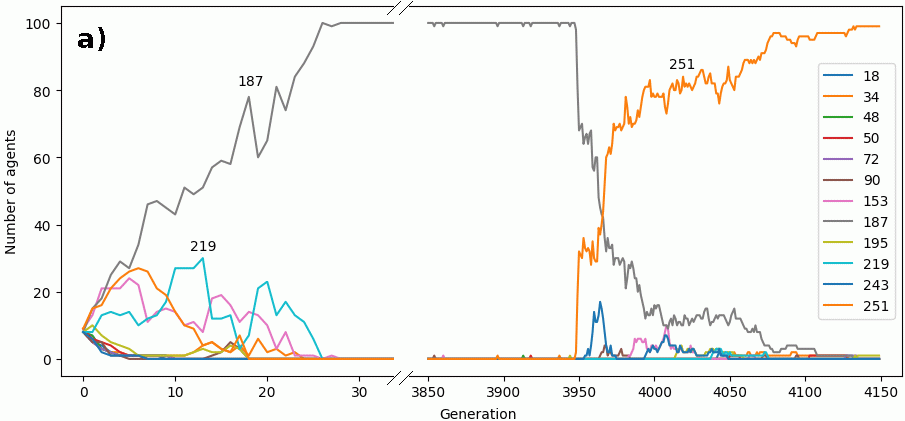}
    
    \includegraphics[width=0.75\linewidth]{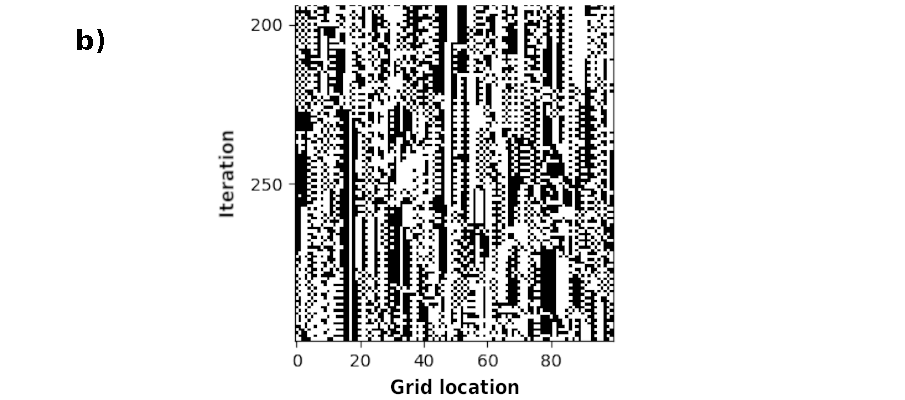}
    \caption{a) An example of long term evolution of a population of 100 agents with randomly assigned strategies over 5,000 generations in baseline scenario (i.e., no noise and no agent mobility. b) Reputation history of grid locations for directed movement enabled and swap=2 }
    \label{fig:baseline_evolution}
\end{figure}

We simulated a population of 100 agents with randomly assigned strategies over 5,000 generations, where each generation lasts 300 iterations and the mutation probability is set to $p_m$ = 0.001. In the baseline scenario-i.e., no noise and no agent mobility (Fig. \ref{fig:baseline_evolution}a), we observe that rule 187 typically dominates the population within the first few dozen generations. However, it is eventually replaced by the most cooperation-promoting rule, rule 251. The timing of this transition varies depending on the random initial conditions and can take up to several thousand generations. 

When agent mobility is introduced via swapping, more frequent short-lived oscillations emerge in the number of agents following dominating rules. Nevertheless, highly donation-prone strategies such as rules 187 and 251 continue to dominate over time. Interestingly, when directed movement is enabled, even the minimal swap value of 1 leads to a highly diverse population in which all rules are represented more equally, but still grid view shows structured pattern (Fig \ref{fig:baseline_evolution}b).

The dominance of donation-maximizing rules can be mitigated by introducing a \textbf{fatigue factor}, which prevents an agent from donating in iteration $i$ if it has donated in each of the previous $n$ iterations. This aligns with principles of motivated learning, in which agents operate in environments that require the management of limited resources to satisfy both primitive and abstract needs. For example, a robotic agent's primitive need might be maintaining battery charge, while an abstract need could involve ensuring access to spare batteries or power banks \cite{Starzyk2010,Starzyk2017,Kowalik_2023}.

\begin{figure}
\centering
    \includegraphics[width=0.65\linewidth]{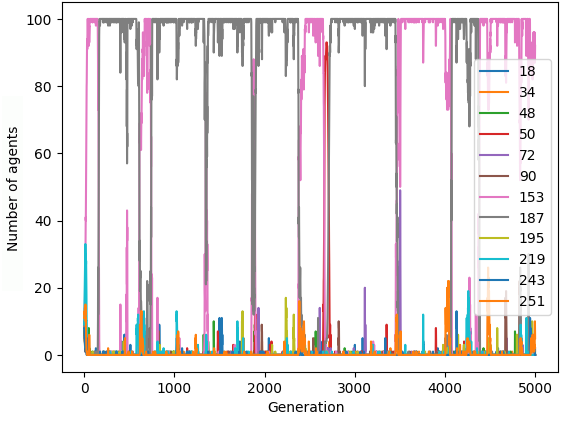}
    \caption{Long-term evolution of a population of 100 agents with randomly assigned strategies over 5,000 generations. The introduction of a fatigue mechanism results in oscillations in the dominance of cooperation strategies (rules).}

    \label{fig_fatigue_factor_oscylations}
\end{figure}

Introducing such a fatigue mechanism reduces the long-term dominance of Rules 251 and 187. The resulting population dynamics resemble the cyclical cooperation and defection patterns (Fig. \ref{fig_fatigue_factor_oscylations}) observed in Nowak and Sigmund's seminal work~\cite{Nowak_Sigmund_1998}.

\section{Discussion, Future Work and Conclusions}

Agent mobility plays an important role in the dynamics of cooperation and the resulting average payoff per agent. Donations can arise through direct reciprocity (repeated interactions between the same agents) or indirect reciprocity (interactions between randomly paired agents who may meet only once per generation).

Recent studies have focused on variations of the Donation Game environment—such as the impact of perception and action noise, or the use of more sophisticated agent strategies like forgiveness and generosity\cite{2023_Griffiths_Oren_Generosity}—primarily under the assumption of fully random interactions, similar to collisions between particles in a gas. We argue that this approach should be extended to incorporate more structured interaction topologies, akin to different "phases of matter," where a set portion of interactions may be local rather than random.
\begin{figure}
    \centering
    \includegraphics[width=0.65\linewidth]{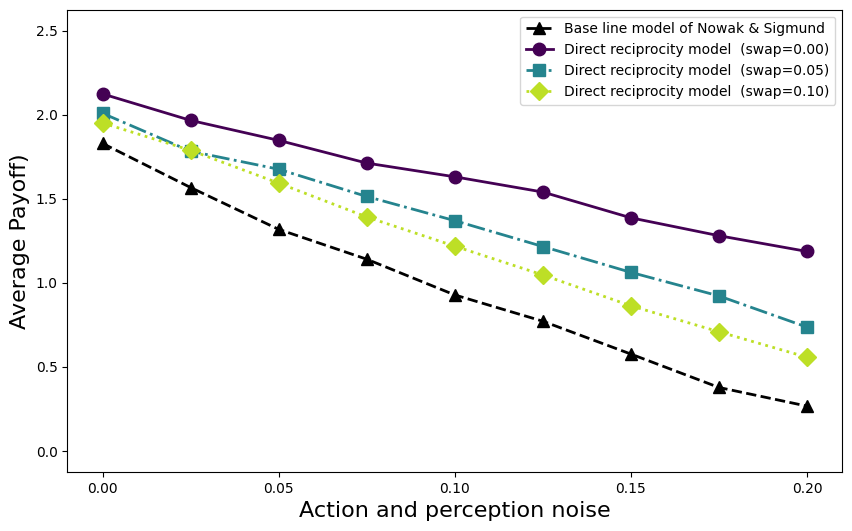}
    
    \caption{Impact of action and perception noise at two swapping levels in a modified version of the Nowak and Sigmund ~\cite{2023_Griffiths_Oren_Generosity, Nowak_Sigmund_1998} model, where cooperation occurs exclusively between nearest neighbors.}
    \label{fig:Mateusz_wykres}
\end{figure}

To support this argument, we adapted the nuanced reputation Donation Game framework proposed by Griffiths and Oren\cite{2023_Griffiths_Oren_Generosity} and based on Nowak and Sigmund model\cite{Nowak_Sigmund_1998} by replacing random partner selection with pairings among nearest neighbors (left or right), while also incorporating random swapping of agent pairs. As shown in Fig. \ref{fig:Mateusz_wykres}, local interactions alone can increase the average payoff by approximately 15\% compared to a noise-free baseline with random interactions. This difference increases when action and perception noises are added. For $a_p$ and $a_e$ set to 0.20, this improvement grows to nearly 300\%.


This paper presents several contributions to the study of reciprocity. First, we introduce a novel framework for modeling cooperation in the Donation Game using Binary Cellular Automata. Unlike traditional models that assume fully random interactions, the BCA approach incorporates spatial locality and enables structured, rule-based agent interactions.

We curate and analyze a set of Wolfram BCA rules, each corresponding to socially interpretable cooperation strategies—such as In-Group Bias, Rank-Based Assistance, and Feudal Systems and Hesitation Dilemma —aligned with Donation Game dynamics. We also introduce a swap parameter to control the ratio of local to non-local interactions, effectively modeling agent mobility and the balance between direct and indirect reciprocity. Our results show that moderate agent mobility fosters emergent behaviors, such as spatial-temporary localized clusters of cooperation or defection, and enables the ordered transmission of reputation between agents. However, high mobility gradually erodes these structures.

Agent mobility proves to be a critical factor: even minimal swapping can reshape the cooperative landscape, triggering new patterns of cooperation or anti-cooperation. Similar effects can be induced by noise, which we find to be asymmetrical in impact. While perception noise may disrupt cooperation under donation-promoting rules, it can paradoxically enhance cooperation under otherwise non-cooperative strategies.

We further demonstrate that each behavioral rule exhibits a unique sensitivity to environmental conditions. For example, Rule 50 shows delayed degradation under action noise, revealing complex nonlinear dynamics. To explore adaptive evolution, we define a mutation matrix between rules and show that, over time, cooperative strategies like Rules 251 and 187 tend to dominate—even when the initial rule distribution is random. However, this dominance can be moderated through the introduction of a fatigue factor, which prevents agents from donating continuously and addresses a fundamental limitation of the classic Donation Game: the assumption of unlimited reputation as a resource.

In future work, we aim to investigate how Donation Game dynamics evolve when agents must manage and share limited resources. This will enable integration with motivated learning frameworks, where cooperation is shaped by agents' internal needs and resource constraints. 

\section{Contributions}
Marcin Kowalik - conceptualization, methodology, experimental design, simulation coding, data analysis, original draft preparation, and project administration. Przemysław Stokłosa - code validation and refactoring. Mateusz Grabowski - simulations related to the adjacent-reputation Donation Game (Fig. \ref{fig:Mateusz_wykres}). Janusz Starzyk and Paweł Raif - manuscript review.

\bibliographystyle{unsrt}  
\bibliography{references}  






\end{document}